\documentstyle[12pt]{article}
\newcommand\C{{\bf C}}
\newcommand\R{{\bf R}}
\newcommand\disp{\displaystyle}
\newcommand\dfrac{\displaystyle\frac}

\begin{document}

\begin{center}
{\large \bf Nonlinear superposition formulas based on
Lie group ${\rm SO}(n+1,n)$}
\end{center}

\vskip0.6cm

{\bf \v{C}. Burd\'{\i}k}

{\it Department of Mathematics and Doppler Institute, FNSPE, Czech
Technical

University,
Trojanova 13, CZ-120 00 Prague 2, Czech Republic}

{\bf O. Navr\'{a}til}

{\it Department of Mathematics, FTS, Czech Technical University,

Na Florenci 25, CZ-110 00 Prague 1, Czech Republic}

\vskip0.6cm

Systems of nonlinear ordinary differential equations are
constructed, for which the general solution is algebraically
expressed in terms of a finite number of particular solutions.
Expressions of that type are called the nonlinear superposition
formulas. These systems are connected with local Lie groups
transformations on their homogeneous spaces. In the presented
work the nonlinear superposition formulas are constructed for the
case of the ${\rm SO}(3,2)$ group and some aspects in the general
case of ${\rm SO}(n+1,n)$ are studied.

\vskip12pt

\begin{center}
{\bf 1. Introduction}
\end{center}

Let us consider a system of $n$ first--order differential equations
\begin{equation}
\label{sz1}
\dot{x}^{\mu}(t)=\chi^{\mu}(x^1,x^2,\dots,x^n,t)\,,\quad
\mu=1,\dots,n\,,
\end{equation}
where the dot denotes differentiation $x^{\mu}(t)$ with respect
to time $t$. It is known for a very long time that, in some cases
which we will specify later, it is possible to express
the general solution as a nonlinear function of a finite number of
particular solutions; it is of the form
\begin{equation}
\label{sz2}
{\bf x}={\bf F}({\bf x}_1,{\bf x}_2,\dots,{\bf x}_m,c_1,\,\dots,c_s)\,,
\qquad{\bf x}\in\R^n\,,
\end{equation}
where ${\bf x}_1$, \dots, ${\bf x}_m$ are particular solutions
(\ref{sz1}), $c_1$, $c_2$, \dots, $c_s$ are arbitrary constants, and
${\bf x}(t)$ is the general solution. These relations are called nonlinear
superposition formulas. Here ${\bf x}$ means a vector with elements
$x^1$, $x^2$, \dots, $x^n$.

An example of such systems is a homogeneous system of the first
order linear differential equations, in which the general solution
is expressed as a linear combination of $n$ linearly independent
particular solutions. The other known example is the Riccati
equation
\begin{equation}
\label{ric}
\dot{x}=a(t)+b(t)x+c(t)x^2\,,
\end{equation}
where $a(t)$, $b(t)$, $c(t)$ are continuous differentiable
functions with respect to $t$. In this case, for any four solutions
$x_i(t)$, $i=1,\dots,4$, the relation
\begin{equation}
\label{ric2}
\frac{x_1(t)-x_3(t)}{x_1(t)-x_4(t)}\cdot
\frac{x_2(t)-x_4(t)}{x_2(t)-x_3(t)}=
\frac{u_1-u_3}{u_1-u_4}\cdot\frac{u_2-u_4}{u_2-u_3}\,,
\end{equation}
where $x_i(0)=u_i$ are initial conditions, is valid.

In the general case, such systems of differential equations are
connected with the local Lie group $G$ of transformations on a
factor space $M=G/G_0$, where $G_0$ is a Lie subalgebra of $G$
\cite{PhysD}. We recall this connection briefly.

By the local Lie group $G$ of transformations on the $M$, we
understand a smooth mapping $\varphi:G\times M\to M$, (we use the
abbreviation $\varphi(g,u)=g\cdot u$), for which: a) $e\cdot
u=u$, for any $u\in M$, where $e$ is the unit element of the
group $G$, b) for any two elements $g_1$, $g_2\in G$ and any
$u\in M$ is $g_2\cdot\bigl(g_1\cdot
u\bigr)=\bigl(g_2g_1\bigr)\cdot u$ and c) $g\cdot u=u$ for any
$u\in M$ imply $g=e$ \cite{Pont}.

In the local coordinate system, we write $x=g\cdot u$ as
\begin{equation}
\label{pon1}
x^{\mu}=f^{\mu}(a^1,\dots,a^N,u^1,\dots,u^n)\,,
\quad\mu=1,\dots,n\,,
\end{equation}
where $N$ is a dimension of the group $G$ and $a^r$, $r=1,\dots,N$,
are their local coordinates. For $x^\mu({\bf a},{\bf u})$, we can
write
\begin{equation}
\label{pon2}
\frac{\partial x^\mu}{\partial a^r}=
\sum_{s=1}^N\xi_s^{\mu}({\bf x})v_r^s({\bf a})\,,
\end{equation}
where the vector fields $X_s({\bf x})=
-\disp\sum_{\mu=1}^n\xi_s^\mu({\bf x})
\frac{\partial}{\partial x^\mu}$ obey the equations
$$
\bigl[X_r,X_s\bigr]=\sum_{t=1}^Nc_{rs}^tX_t\,,
$$
in which $c_{rs}^t$ are structure constants of the Lie algebra
of the group $G$.

Conversely, the vector fields $X_k({\bf x})$ determine the
infinitesimal action of the local Lie group on the space $M$
uniquely \cite{Pont}.

Let $g(t)=\bigl(a^1(t),\dots,a^N(t)\bigr)$, $t\in\R$, be a curve
in the Lie group $G$ such that $g(0)=e$. This gives a curve in
the space $M$. Derivation of equation (\ref{pon1}) with respect
to the parameter $t$ gives, by using (\ref{pon2}), the system
of differential equations
\begin{equation}
\label{pon3}
\dot{x}^\mu=\sum_{r=1}^N \xi_r^\mu({\bf x})Z^r(t)\,,
\quad \mu=1,\dots,n\,.
\end{equation}
In our paper we will deal with such systems connected with the
Lie group ${\rm SO}(n+1,n)$.

If the system of equations (\ref{sz1}) has form (\ref{pon3}), then
there is a curve in some local coordinates on the Lie group $G$,
which acts on the factor space $M$. In this case, it is possible to
find the superposition formula \cite{PhysD}. Any particular solution
of the system (\ref{pon3}) can be written in the form
\begin{equation}
{\bf x}_k(t)=g(t)\cdot{\bf u}_k\,,
\end{equation}
where ${\bf u}_k={\bf x}_k(0)$ is the initial condition.

We express the action of the local group $G$ by using
action of this group to few points of the space $M$, which is
supposed known. In principle, it means to find from the system
of the equations
\begin{equation}
\label{pon4}
\begin{array}{l}
{\bf x}_1={\bf f}\bigl(a^1,\dots,a^N,{\bf u}_1\bigr)\\
{\bf x}_2={\bf f}\bigl(a^1,\dots,a^N,{\bf u}_2\bigr)\\
\qquad\qquad\dots\\
{\bf x}_r={\bf f}\bigl(a^1,\dots,a^N,{\bf u}_r\bigr)
\end{array}
\end{equation}
the coordinates of the group $a^i$.
To be able to find the group coordinates $a^i$, $i=1,\dots,N$,
we should use the action of $r$ points. It is evident that the
number $r$ must fulfil the inequality $nr\geq N$, where $n$ is
the dimension of $M$, and $N$ is the dimension of the group $G$.

If we solve this, then we are able to express the elements of Lie
group $G$ by means of the known transformations of $r$ points
in the form
\begin{equation}
g=g\bigl({\bf x}_1,{\bf x}_2,\dots,{\bf x}_r,
{\bf u}_1,{\bf u}_2,\dots,{\bf u}_r\bigr)\,,
\end{equation}
therefore
\begin{equation}
\label{pon5}
{\bf x}=g\bigl({\bf x}_1,\dots,{\bf x}_r,
{\bf u}_1,\dots,{\bf u}_r\bigr)\cdot{\bf u}
\end{equation}
holds. Now we see that formula (\ref{pon5}) is invariant with respect
to the action of the local group $G$.

If ${\bf x}_i(t)$, $i=1,\dots,r$, are known solutions of the system
(\ref{pon3}) for given functions $Z^r(t)$, then any other solution
${\bf x}(t)$ of that system is given by (\ref{pon5}). Therefore the
relation (\ref{pon5}) is the superposition formula \cite{PhysD}.

For example, the Riccati equation (\ref{ric}) is connected with the
Lie group $G=SL(2)$ that acts on the space $M$ as
$$
g(t)\cdot u=x(t)=\frac{a_{21}(t)+ua_{22}(t)}{a_{11}(t)+ua_{12}(t)}\,,
$$
where we represented the elements of the group $G$ by matrix
$\left(\begin{array}{cc}a_{11}&a_{12}\\
a_{21}&a_{22}\end{array}\right)$, with the determinant equals to $1$.

In the eighties, there where a lot of papers \cite{JMP1}--\cite{JMP4}
which were devoted to the systems of differential equations of
this type and especially to finding the superposition formulas for
these systems. For the most part, the authors study the systems,
which are connected with Lie groups ${\rm SL}(n,\R)$ or
${\rm SL}(n,\C)$. In the
general paper \cite{PhysD} the methods are described that are
used for constructing the superposition formulas. They are
demonstrated with examples, in which the Lie groups ${\rm SL}(n,\R)$
and ${\rm O}(p+1,n-p+1)$ act on simple projective spaces.

In \cite{JMP1} the more general case of the projective matrix
Riccati equation are studied for ${\rm SL}(2n,\R)$ and ${\rm
SP}(2n,\R)$. In paper \cite{JMP2}, the systems that arise from
the action of ${\rm SL}(n,\C)$ on the factor spaces ${\rm
SL}(n,\C)/{\rm O}(n,\C)$ and ${\rm SL}(2n,\C)/{\rm SP}(2n,\C)$ are
studied. Paper \cite{JMP3}, is devoted to the superposition
formulas for rectangular--matrix Riccati equations on the space
${\rm SL}(n+k,\C)/P(k)$, where $P(k)$ are special maximal
parabolic subgroups of the ${\rm SL}(n+k,\C)$. In \cite{JMP5},
the authors deal with systems connected with the Lie group ${\rm
SU}(n,n)$, and in \cite{JMP6}, th same method for ${\rm SO}(n,n)$
is used. Finally, in paper \cite{JMP4}, the authors study the
systems of equations that are connected with the action of Lie
group ${\rm SL}(n,\C)$ on the space $M={\rm SL}(n,\C)/G_0$, where
$G_0$ is a special non--maximal parabolic subgroup.

The authors mostly used a set of special solutions for reconstructing
the group action on the space $M$. This approach simplifies the
solution of (\ref{pon4}). But on the other hand, we are not able to
use the resulting superposition formulas directly for constructing
the solution of the system on the basis any set of particular
solutions. By means of the action of groups elements, we must first
transform our particular solutions to a special set of particular
solutions used in superposition formulas.

In the next, we study the systems of equations (\ref{pon3})
which are connected with the action of Lie group
${\rm SO}(n+1,n)$ on the space $M={\rm SO}(n+1,n)/P$, where $P$
is one of the maximal parabolic subgroups. To our knowledge
systems of this type were not studied so far. In distinction to
the papers cited above, we do not choose a special set of
particular solutions for reconstructing the group action on
$M$, or, in the terminology of paper \cite{PhysD}, we will
construct group invariants which give nonlinear superposition
formulas in implicit form.

\vskip0.6cm

\begin{center}
{\bf 2. Lie group ${\rm SO}(n+1,n)$ and its Lie algebra}
\end{center}

\vskip0.6cm

In this section, we fix the notation. The Lie group
${\rm SO}(n+1,n)$ is a group of real matrices ${\bf G}$ with
dimension $(2n+1)\times(2n+1)$ that fulfil the equations
\begin{equation}
\label{gr1}
{\bf G}^T\cdot\sigma\cdot{\bf G}=\sigma\,,\qquad
\mbox{where~}\,
\sigma=\left(\begin{array}{ccc}
1&0&0\\0&0&{\bf I}\\0&{\bf I}&0\end{array}\right),
\end{equation}
${\bf G}^T$ denotes a transposed matrix and ${\bf I}$ is a unit
matrix with dimension $n\times n$.

Matrix ${\bf G}$ is written in the form
\begin{equation}
\label{gr2}
{\bf G}=\left(\begin{array}{ccc}
g_{11}&{\bf g}_{12}^T&{\bf g}_{13}^T\\
{\bf g}_{21}&{\bf G}_{22}&{\bf G}_{23}\\
{\bf g}_{31}&{\bf G}_{32}&{\bf G}_{33}\end{array}\right),
\end{equation}
where ${\bf g}$ denotes a column vector, and ${\bf g}^T$ is its
transposed vector represented as row. Therefore ${\bf
g}{\bf g}^T$ is an $n\times n$ matrix, and ${\bf g}^T{\bf g}$ is an
inner product.

If we insert (\ref{gr2}) into (\ref{gr1}) we obtain the following
formulas for elements of the matrix ${\bf G}$
\begin{equation}
\label{gr0}
\begin{array}{l}
g_{11}^2+{\bf g}_{21}^T{\bf g}_{31}+{\bf g}_{31}^T{\bf g}_{21}=1\\
\noalign{\vskip3pt}
g_{11}{\bf g}_{12}+{\bf G}_{22}^T{\bf g}_{31}+
{\bf G}_{32}^T{\bf g}_{21}=0\\
\noalign{\vskip3pt}
g_{11}{\bf g}_{13}+{\bf G}_{23}^T{\bf g}_{31}+
{\bf G}_{33}^T{\bf g}_{21}=0\\
\noalign{\vskip3pt}
{\bf g}_{12}{\bf g}_{12}^T+{\bf G}_{22}^T{\bf G}_{32}+
{\bf G}_{32}^T{\bf G}_{22}=0\\
\noalign{\vskip3pt}
{\bf g}_{13}{\bf g}_{13}^T+{\bf G}_{23}^T{\bf G}_{33}+
{\bf G}_{33}^T{\bf G}_{23}=0\\
\noalign{\vskip3pt}
{\bf g}_{12}{\bf g}_{13}^T+{\bf G}_{22}^T{\bf G}_{33}+
{\bf G}_{32}^T{\bf G}_{23}={\bf I}\end{array}
\end{equation}

In this realization the Lie algebra ${\rm so}(n+1,n)$ is given
by real matrices
$$
{\bf A}=\left(\begin{array}{ccc}
0&{\bf x}^T&{\bf z}^T\\
-{\bf z}&{\bf H}&{\bf W}\\
-{\bf x}&{\bf Y}&-{\bf H}^T\end{array}\right),
$$
where ${\bf W}^T=-{\bf W}$ a ${\bf Y}^T=-{\bf Y}$.

In the group ${\rm SO}(n+1,n)$, we take a subgroup $G_0$ that is
generated by the matrices
$$
{\bf G}_0=\left(\begin{array}{ccc}1&0&0\\0&{\bf D}&0\\
0&0&\bigl({\bf D}^T\bigr)^{-1}\end{array}\right)\cdot
\left(\begin{array}{ccc}1&0&{\bf z}^T\\-{\bf z}&{\bf I}&{\bf Z}\\
0&0&{\bf I}\end{array}\right),
$$
where the equality ${\bf Z}+{\bf Z}^T+{\bf z}{\bf z}^T=0$ holds.
The factor space $M=G/G_0$ can be represented by the matrices
$$
\Xi=\left(\begin{array}{ccc}1&{\bf x}^T&0\\0&{\bf I}&0\\
-{\bf x}&{\bf X}&{\bf I}\end{array}\right),
$$
where ${\bf X}+{\bf X}^T+{\bf x}{\bf x}^T=0$. As coordinates on
$M$, we choose ${\bf x}$ and the antisymmetric part of the matrix
${\bf X}$; this means that
\begin{equation}
\label{dod1}
{\bf Y}={\bf X}+\frac12\,{\bf x}{\bf x}^T\,.
\end{equation}
The action of the group $G$ on the factor space $M$ in these
coordinates can be obtained from the equation
$$
\begin{array}{l}
\left(\begin{array}{ccc}
g_{11}&{\bf g}_{12}^T&{\bf g}_{13}^T\\
{\bf g}_{21}&{\bf G}_{22}&{\bf G}_{23}\\
{\bf g}_{31}&{\bf G}_{32}&{\bf G}_{33}\end{array}\right)\cdot
\left(\begin{array}{ccc}1&{\bf u}^T&0\\0&{\bf I}&0\\
-{\bf u}&{\bf U}&{\bf I}\end{array}\right)=\\
\noalign{\vskip4pt}
\qquad\qquad\qquad
=\left(\begin{array}{ccc}1&{\bf x}^T&0\\0&{\bf I}&0\\
-{\bf x}&{\bf X}&{\bf I}\end{array}\right)\cdot
\left(\begin{array}{ccc}1&0&0\\0&{\bf D}&0\\
0&0&\bigl({\bf D}^T\bigr)^{-1}\end{array}\right)\cdot
\left(\begin{array}{ccc}1&0&{\bf z}^T\\
-{\bf z}&{\bf I}&{\bf Z}\\0&0&{\bf I}\end{array}\right).
\end{array}
$$
If we compare the coefficients on both sides of this equation, we
obtain the following formulas
\begin{equation}
\label{gr3}
\begin{array}{l}
{\bf D}={\bf G}_{22}+{\bf g}_{21}{\bf u}^T+
{\bf G}_{23}{\bf U}\\
\noalign{\vskip3pt}
{\bf x}=\bigl({\bf D}^T\bigr)^{-1}
\bigl({\bf g}_{12}+g_{11}{\bf u}+{\bf U}^T{\bf g}_{13}\bigr)\\
\noalign{\vskip3pt}
{\bf X}=\bigl({\bf G}_{32}+{\bf g}_{31}{\bf u}^T+
{\bf G}_{33}{\bf U}\bigr){\bf D}^{-1}\end{array}
\end{equation}
for the group elements.
Because we restrict ourselves to the local Lie group, we can
suppose that matrix ${\bf D}$ is invertible.

Starting with the action of this group on the space $M$ and by
using the expansion to the first order, we derive an explicit
expression for vector fields, in the basis of the algebra
${\rm so}(n+1,n)$ in the representation
$\bigl(T_{g}f\bigr)(m)=f\bigl(g^{-1}\cdot m)$. Specifically we get
$$
\begin{array}{rcl}
Y_{ij}&\mapsto&-\dfrac{\partial}{\partial Y_{ij}}\\
\noalign{\vskip3pt}
x_i&\mapsto&-\dfrac{\partial}{\partial x_i}-
\dfrac12\sum_{r=1}^nx_r\frac{\partial}{\partial Y_{ri}}\\
\noalign{\vskip3pt}
D_{ij}&\mapsto&x_j\dfrac{\partial}{\partial x_i}-
\sum_{r=1}^nY_{ir}\frac{\partial}{\partial Y_{rj}}\\
\noalign{\vskip3pt}
z_i&\mapsto&-\disp\sum_{r=1}^n\left(Y_{ir}+\frac{x_ix_r}2\right)
\frac{\partial}{\partial x_r}-
\frac12\sum_{r,s=1}^nY_{ri}x_s\frac{\partial}{\partial Y_{rs}}\\
\noalign{\vskip3pt}
W_{ij}&\mapsto&\disp\sum_{r=1}^n\bigl(x_jY_{ri}-x_iY_{rj}\bigr)
\frac{\partial}{\partial x_r}-
\sum_{r,s=1}^nY_{ir}Y_{js}\frac{\partial}{\partial Y_{rs}}\,,
\end{array}
$$
where we define $Y_{rs}=-Y_{sr}$.

The system of differential equations (\ref{pon3}) is in this case
of the form
\begin{equation}
\label{gr4}
\begin{array}{rcl}
\dot{\bf x}(t)&=&{\bf a}(t)-{\bf B}(t){\bf x}-{\bf Y}{\bf c}(t)
+\dfrac12{\bf x}{\bf c}^T(t){\bf x}-{\bf Y}{\bf C}(t){\bf x}\\
\noalign{\vskip3pt}
\dot{\bf Y}(t)&=&{\bf A}(t)+
\dfrac12\,\bigl({\bf x}{\bf a}^T(t)-{\bf a}(t){\bf x}^T\bigr)-
\bigl({\bf Y}{\bf B}(t)+{\bf B}^T(t){\bf Y}\bigr)+\\
\noalign{\vskip3pt}
&&+\dfrac12\,\bigl({\bf Y}{\bf c}(t){\bf x}^T+
{\bf x}{\bf c}^T(t){\bf Y}\bigr)-
{\bf Y}{\bf C}(t){\bf Y}\end{array}
\end{equation}
where ${\bf a}(t)$ and ${\bf c}(t)$ are vector functions
differentiable with respect to $t$; ${\bf A}(t)$, ${\bf B}(t)$
and ${\bf C}(t)$ are differentiable matrix functions, for which
${\bf A}^T=-{\bf A}$, ${\bf C}^T=-{\bf C}$ and ${\bf Y}^T=-{\bf Y}$.

This system of differential equation is the same as in the paper \cite{JMP7},
where is given classification of all systems of nonlinear ordinary differential
equations of this type.

\vskip12pt

\begin{center}
{\bf 3. Representations of the action of the group by means of
particular solutions}
\end{center}

The system of equations (\ref{gr4}) arises from the action of the
Lie group ${\rm SO}(n+1,n)$ on the factor space
$M={\rm SO}(n+1,n)/G_0$. Therefore, one can find the superposition
formula. The action of the group ${\rm SO}(n+1,n)$ on the space $M$
is given by the relations (\ref{gr3}). We try to express this action
by means of any known solutions ${\bf x}_k=g\cdot{\bf u}_k$.
If ${\bf x}_k(t)$ are solutions of the differential equations
(\ref{gr4}) with the initial conditions ${\bf x}_k(0)={\bf u}_k$,
we obtain the general solution ${\bf x}(t)$ with the initial
condition ${\bf x}(0)={\bf u}$ in the form
$$
{\bf x}(t)=G\bigl({\bf x}_1(t),\dots,{\bf x}_r(t),
{\bf u}_1,\dots,{\bf u}_r\bigr)\cdot{\bf u}\,,
$$
where $g\bigl({\bf x}_1(t),\dots,{\bf x}_r(t),
{\bf u}_1,\dots,{\bf u}_r\bigr)=g(t)$ is the expression of an element
of the group $G$ in terms of the known transformation.

Dimension of the group ${\rm SO}(n+1,n)$ is equal to $n(2n+1)$ and the
dimension of the space $M=SO(n+1,n)/G_0$ is $\dfrac{n(n+1)}2$. So,
to express $N$ coordinates of the group elements by using the known
solutions, we must know at least $r$ solutions, where $r$ fulfils
the inequality $n(2n+1)\leq\dfrac{n(n+1)}2\,r$. Consequently,
for $n=1$, it is sufficient to known three particular solutions,
and for $n>1$, we must know at least, four particular solutions of
the system (\ref{gr4}).

By ${\bf x}_i$ we denote a vector, and by ${\bf X}_i={\bf
Y}_i-\dfrac12\,{\bf x}_i{\bf x}_i^T$, a matrix for and $i^{th}$
solution of the system of differential equations (\ref{gr4}) with
initial conditions ${\bf u}_i$ and ${\bf U}_i$; ${\bf D}_i$ is an
invertible matrix defined for an appropriate solution by the
first equation in (\ref{gr3}). Further, we denote ${\bf
u}_{ik}={\bf u}_i-{\bf u}_k$, ${\bf x}_{ik}={\bf x}_i-{\bf x}_k$.
Similarly, we define ${\bf X}_{ik}$ and ${\bf U}_{ik}$. Temporarily
we suppose that all matrices which we will use in our calculations
are invertible.

From equation (\ref{gr3}) we obtain some coordinates of elements
of the group $G$. Readily we discover that for indices $i$ and $k$,
the relations
$$
\begin{array}{l}
{\bf g}_{13}=\bigl({\bf U}_{ik}^T\bigr)^{-1}\cdot
\bigl({\bf D}_i^T{\bf x}_i-{\bf D}_k^T{\bf x}_k-
g_{11}{\bf u}_{ik}\bigr)\\
\noalign{\vskip3pt}
{\bf G}_{23}=\bigl({\bf D}_i-{\bf D}_k-
{\bf g}_{21}{\bf u}_{ik}^T\bigl)\cdot{\bf U}_{ik}^{-1}\\
\noalign{\vskip3pt}
{\bf G}_{33}=\bigl({\bf X}_i{\bf D}_i-{\bf X}_k{\bf D}_k-
{\bf g}_{31}{\bf u}_{ik}^T\bigr)\cdot{\bf U}_{ik}^{-1}\\
\noalign{\vskip3pt}
{\bf g}_{12}={\bf D}_i^T{\bf x}_i-g_{11}{\bf u}_i-
{\bf U}_i^T{\bf g}_{13}\\
\noalign{\vskip3pt}
{\bf G}_{22}={\bf D}_i-{\bf g}_{21}{\bf u}_i^T-
{\bf G}_{23}{\bf U}_i\\
\noalign{\vskip3pt}
{\bf G}_{32}={\bf X}_i{\bf D}_i-{\bf g}_{31}{\bf u}_i^T-
{\bf G}_{33}{\bf U}_i
\end{array}
$$
are valid. If we now apply these equations to (\ref{gr0}), after
a simple algebra we obtain the following formulas
\begin{equation}
\label{res1}
\begin{array}{l}
g_{11}^2+{\bf g}_{21}^T{\bf g}_{31}+{\bf g}_{31}^T{\bf g}_{21}=1\\
\noalign{\vskip3pt}
{\bf D}_i^T\cdot\bigl(g_{11}{\bf x}_i+{\bf g}_{31}+
{\bf X}_i^T{\bf g}_{21}\bigr)={\bf u}_i\\
\noalign{\vskip3pt}
{\bf D}_i^T\cdot\bigl({\bf x}_i{\bf x}_k^T+
{\bf X}_i^T+{\bf X}_k\bigr)\cdot{\bf D}_k=
{\bf u}_i{\bf u}_k^T+{\bf U}_i^T+{\bf U}_k
\end{array}
\end{equation}
From the second equation in (\ref{res1}) we can express
\begin{equation}
\label{res2}
\begin{array}{l}
{\bf g}_{21}=-\bigl({\bf X}_{ik}^T\bigr)^{-1}
\Bigl[g_{11}{\bf x}_i-\bigl({\bf D}_i^T\bigr)^{-1}{\bf u}_i\Bigr]-
\bigl({\bf X}_{ki}^T\bigr)^{-1}
\Bigl[g_{11}{\bf x}_k-\bigl({\bf D}_k^T\bigr)^{-1}{\bf u}_k\Bigr]\\
\noalign{\vskip3pt}
{\bf g}_{31}={\bf X}_k^T\bigl({\bf X}_{ik}^T\bigr)^{-1}
\Bigl[g_{11}{\bf x}_i-\bigl({\bf D}_i^T\bigr)^{-1}{\bf u}_i\Bigr]+
{\bf X}_i^T\bigl({\bf X}_{ki}^T\bigr)^{-1}
\Bigl[g_{11}{\bf x}_k-\bigl({\bf D}_k^T\bigr)^{-1}{\bf u}_k\Bigr]
\end{array}
\end{equation}
Next we use the notation
\begin{equation}
\label{res3}
\begin{array}{l}
\Omega_{ik}={\bf x}_i{\bf x}_k^T+{\bf X}_i^T+{\bf X}_k\\
\noalign{\vskip3pt}
\omega_{ik}={\bf u}_i{\bf u}_k^T+{\bf U}_i^T+{\bf U}_k\\
\noalign{\vskip3pt}
{\bf h}_i=g_{11}{\bf x}_i-\bigl({\bf D}_i^T\bigr)^{-1}{\bf u}_i
\end{array}
\end{equation}
It is easy to see that the relations
$$
\begin{array}{lcl}
{\bf x}_i{\bf x}_k^T+\Omega_{ik}+\Omega_{ki}=0
\quad&\mbox{and}&\quad
\Omega_{ik}^T=\Omega_{ki}\\
\noalign{\vskip3pt}
{\bf u}_i{\bf u}_k^T+\omega_{ik}+\omega_{ki}=0
\quad&\mbox{and}&\quad
\omega_{ik}^T=\omega_{ki}
\end{array}
$$
are valid. By simple algebraic calculations from equations
(\ref{res1}) and (\ref{res2}) for any $i$, $j$, and $k$,
we derive the following formulas
\begin{eqnarray}
\label{res4.1}
&&{\bf D}_i^T\Omega_{ik}{\bf D}_k=\omega_{ik}\\
\noalign{\vskip3pt}
\label{res4.2}
&&{\bf h}_k+
{\bf X}_{jk}^T\cdot\bigl({\bf X}_{ij}^T\bigr)^{-1}{\bf h}_i+
{\bf X}_{ik}^T\cdot\bigl({\bf X}_{ji}^T\bigr)^{-1}{\bf h}_j=0\\
\noalign{\vskip3pt}
\label{res4.3}
&&\begin{array}{l}
g_{11}^2+
\Bigl[{\bf h}_k^T{\bf X}_{ik}^{-1}{\bf x}_k+
{\bf h}_k^T{\bf X}{_ki}^{-1}{\bf x}_i\Bigr]\cdot
\Bigl[{\bf x}_k^T\bigl({\bf X}_{ik}^T\bigr)^{-1}{\bf h}_i+
{\bf x}_i^T\bigl({\bf X}_{ki}^T\bigr)^{-1}{\bf h}_k\Bigr]-\\
\noalign{\vskip2pt}
\qquad
-{\bf h}_i^T{\bf X}_{ik}^{-1}\Omega_{ik}
\bigl({\bf X}_{ki}^T\bigr)^{-1}{\bf h}_k-
{\bf h}_k^T{\bf X}_{ki}^{-1}\Omega_{ki}
\bigl({\bf X}_{ik}^T\bigr)^{-1}{\bf h}_i=1\,.
\end{array}
\end{eqnarray}
We found the system of equations from which it is possible to
determine the matrix (\ref{gr2}). Since we restricted ourselves
to the neighbourhood of the point $t=0$, we can suppose that
matrices ${\bf D}_i(t)$ are invertible. From the relation
(\ref{res4.1}) it follows that in this neighbourhood matrices
$\Omega_{ik}(k)$ and $\omega_{ki}$ are simultaneously invertible
or non--invertible. So, it is easy to see that, if matrix
$\omega_{ik}$ is invertible, it is enough to know only one
solution of this system ${\bf D}_i$ for one value of $i$. We can
find other ${\bf D}_k$ from equation (\ref{res4.1}). If we know
${\bf D}_i$ explicitly, the equation (\ref{res4.3}) is a
quadratic equation with respect to $g_{11}$. The odd equations
can then give the matrices ${\bf D}_i$.

We enunciate the previous account in terms of the following theorem.

\vskip0.6cm

\noindent {\bf Theorem:} Let ${\bf x}_i(t)$ and ${\bf Y}_i(t)$,
$i=1,\,2,\,3$, be three solutions of the equation (\ref{gr4}),
${\bf X}_i(t)={\bf Y}_i(t)-\dfrac12\,{\bf x}_i{\bf x}_i^T$,
${\bf u}_i={\bf x}_i(0)$, ${\bf U}_i={\bf X}_i(0)$ and let the
matrices ${\bf U}_{ik}$ be invertible. Then, there is a
neighbourhood of the point $t=0$, matrices ${\bf D}_i(t)$,
$i=1,\,2,\,3$, ${\bf D}_i(0)={\bf I}$, and function
$g_{11}(t)$, $g_{11}(0)=1$ such, that, for them formulas
(\ref{res4.1}), (\ref{res4.2}), and (\ref{res4.3}) are true.

\vskip0.4cm

\begin{center}
{\bf 4. The superposition formulas}
\end{center}

In the previous sections, we formulated conditions
(\ref{res4.1})--(\ref{res4.3}) that are valid for any three
solutions of the system (\ref{gr4}). Although we are not able to
solve the system (\ref{res4.1})--(\ref{res4.3}) explicitly, it is
possible to derive, from them, certain relations for solutions of
the system of differential equations.

We suppose that we have five solutions of this system, and there
exists the inverse matrix to $\omega_{ik}$ for any
$i\neq k$, $i,\,k=1,\,\dots,\,5$. Then from the equation
(\ref{res4.1}), we obtain
$$
{\bf D}_k^{-1}\Omega_{rk}^{-1}\bigl({\bf D}_r^T\bigr)^{-1}=
\omega_{rk}^{-1}\,.
$$
If we now multiply the equation (\ref{res4.1}) by this equation
from the left, and then, by (\ref{res4.1}) from the left for the
couple $(ri)$, we obtain equation
\begin{equation}
\label{sf1}
{\bf D}_i^T\Omega_{ik}\Omega_{rk}^{-1}\Omega_{ri}{\bf D}_i=
\omega_{ik}\omega_{rk}^{-1}\omega_{ri}\,,
\end{equation}
that is true for any $i$, $k$, and $r$. We multiply this equation
further from the right by the inverse equation (\ref{sf1}) for the
ternary $(sti)$. Then, we obtain
$$
{\bf D}_i^{-1}\Omega_{ti}^{-1}\Omega_{ts}\Omega_{is}^{-1}
\Omega_{ik}\Omega_{rk}^{-1}\Omega_{ri}{\bf D}_i=
\omega_{ti}^{-1}\omega_{ts}\omega_{is}^{-1}
\omega_{ik}\omega_{rk}^{-1}\omega_{ri}\,,
$$
which is true for any $(ikrst)$. If now we put $s=k$ in this
equation, we obtain, for any four solutions of (\ref{gr4}),
the relation
\begin{equation}
\label{sf2}
{\bf D}_i^{-1}\Omega_{si}^{-1}\Omega_{sk}
\Omega_{rk}^{-1}\Omega_{ri}{\bf D}_i=
\omega_{si}^{-1}\omega_{sk}\omega_{rk}^{-1}\omega_{ri}\,.
\end{equation}
It means that matrices
$\Omega_{si}^{-1}\Omega_{sk}\Omega_{rk}^{-1}\Omega_{ri}$ and
$\omega_{si}^{-1}\omega_{sk}\omega_{rk}^{-1}\omega_{ri}$ are
similar. Therefore, all their invariants are identical.

Now, we will use this interesting property of the solution of
differential equations (\ref{gr4}) to obtain superposition
formulas of the system (\ref{gr4}) for small $n$.

\vskip9pt

For $n=1$ we get the Lie group ${\rm SO}(2,1)$. In this case,
the vector ${\bf x}$ is reduced to the number $x$, and ${\bf Y}=0$.
From this we have ${\bf X}_i=-\dfrac12\,x_i^2$ and
$\Omega_{ik}=-\dfrac12\,(x_i-x_k)^2$. In this case, equation
(\ref{sf2}) has after the extraction form
$$
\frac{x_s-x_k}{x_s-x_i}\cdot\frac{x_r-x_i}{x_r-x_k}=
\frac{u_s-u_k}{x_u-u_i}\cdot\frac{u_r-u_i}{u_r-u_k}\,.
$$
That represents the well--known superposition formula (\ref{ric2})
for the Riccati equation (\ref{ric}). This is a consequence of the
isomorphism between ${\rm so}(2,1)$ and ${\rm sl}(2)$.

\vskip9pt

Now, we will study the case $n=2$, that is the Lie group
${\rm SO}(3,2)$. In this case, we have
$$
{\bf Y}=\left(\begin{array}{cc}0&x_3\\-x_3&0\end{array}\right)
\quad\mbox{and}\quad
{\bf X}=\frac12\left(\begin{array}{cc}
-x_1^2&-x_1x_2+2x_3\\-x_1x_2-2x_3&-x_2^2\end{array}\right).
$$
If we denote the components of the $i^{th}$ solution by
$x_1^{(i)}$, $x_2^{(i)}$ a $x_3^{(i)}$, we obtain
$$
\det\bigl({\bf U}_{ik}\bigr)=\frac14\,
\Bigl(4\bigl(u_3^{(i)}-u_3^{(k)}\bigr)^2-
\bigl(u_2^{(i)}u_1^{(k)}-u_1^{(i)}u_2^{(k)}\bigr)^2\Bigr)
$$
and the determinants of the matrices $\Omega_{ik}$ and
$\omega_{ik}$ are
\begin{equation}
\label{sf3}
\det\bigl(\Omega_{ik}\bigr)=\frac14\,\Delta_{ik}^2
\quad\mbox{and}\quad
\det\bigl(\omega_{ik}\bigr)=\frac14\,\delta_{ik}^2\,,
\end{equation}
where
\begin{equation}
\label{sf4}
\Delta_{ik}=2x_3^{(i)}-2x_3^{(k)}+
x_2^{(i)}x_1^{(k)}-x_1^{(i)}x_2^{(k)}
\quad\mbox{and}\quad
\delta_{ik}=2u_3^{(i)}-2u_3^{(k)}+
u_2^{(i)}u_1^{(k)}-u_1^{(i)}u_2^{(k)}\,.
\end{equation}
We see that the conditions for the matrices ${\bf U}_{ik}$ to be
invertible imply the invertibility for matrices $\omega_{ik}$.

If we now take the determinant in equation (\ref{sf2}) we obtain
for any four different solutions the equality
\begin{equation}
\label{sf5}
\frac{\Delta_{sk}}{\Delta_{si}}\cdot
\frac{\Delta_{ri}}{\Delta_{rk}}=
\frac{\delta_{sk}}{\delta_{si}}\cdot
\frac{\delta_{ri}}{\delta_{rk}}\,.
\end{equation}
As it was mentioned above, in this case, we can construct the general
solution of the system (\ref{gr4}) by using four particular
solutions. Take now five solutions, for which from (\ref{sf5})
we obtain independent equations
\begin{eqnarray}
\label{sf6.4}
&\disp\frac{\Delta_{24}}{\Delta_{14}}\cdot
\frac{\Delta_{13}}{\Delta_{23}}=
\frac{\delta_{24}}{\delta_{14}}\cdot
\frac{\delta_{13}}{\delta_{23}}\,,\qquad
\frac{\Delta_{34}}{\Delta_{14}}\cdot
\frac{\Delta_{12}}{\Delta_{23}}=
\frac{\delta_{34}}{\delta_{14}}\cdot
\frac{\delta_{12}}{\delta_{23}}\\
\noalign{\vskip3pt}
\label{sf6.5}
&\disp\frac{\Delta_{25}}{\Delta_{15}}\cdot
\frac{\Delta_{13}}{\Delta_{23}}=
\frac{\delta_{25}}{\delta_{15}}\cdot
\frac{\delta_{13}}{\delta_{23}}\,,\qquad
\frac{\Delta_{35}}{\Delta_{15}}\cdot
\frac{\Delta_{12}}{\Delta_{23}}=
\frac{\delta_{35}}{\delta_{15}}\cdot
\frac{\delta_{12}}{\delta_{23}}\,,\qquad
\frac{\Delta_{45}}{\Delta_{15}}\cdot
\frac{\Delta_{12}}{\Delta_{24}}=
\frac{\delta_{45}}{\delta_{15}}\cdot
\frac{\delta_{12}}{\delta_{24}}\,.
\end{eqnarray}

Equations (\ref{sf6.5}) are understood as a system of linear
equations for $x_1^{(5)}$, $x_2^{(5)}$, and $x_3^{(5)}$. This
system has a solution when its determinant $D(t)$ is different
from zero. By direct calculation for $t=0$, we obtain
\begin{equation}
\label{sf7}
D(0)=\bigl(\delta_{15}-\delta_{25}\bigr)
\bigl(\delta_{13}-\delta_{14}+\delta_{34}\bigr)-
\bigl(\delta_{15}-\delta_{35}\bigr)
\bigl(\delta_{12}-\delta_{14}+\delta_{24}\bigr)+
\bigl(\delta_{15}-\delta_{45}\bigr)
\bigl(\delta_{12}-\delta_{13}+\delta_{23}\bigr)\,.
\end{equation}
The terms in the second parentheses do not depend on $u_3^{(i)}$
and are not identical zero. Because the terms in first parentheses
depend on $u_3^{(i)}-u_3^{(k)}$, this determinant is not identical
zero for any possible initial conditions ${\bf u}_i$ and ${\bf U}_i$.
As the determinant $D(t)$ is given by the solutions of system
(\ref{gr4}) which are continuous and $D(0)\neq0$, there is
any neighbourhood of $t=0$, in which the determinant $D(t)\neq0$.
We see that, in this neighbourhood we can determine, from the system
of the equations (\ref{sf6.5}), the solutions $x_1^{(5)}$, $x_2^{(5)}$
and $x_3^{(5)}$ of the system of differential equations (\ref{gr4}) by
using the particular solution ${\bf x}_i$, ${\bf Y}_i$ for
$i=1,\dots,4$. In other words, formulas (\ref{sf5}) give the implicit
form of nonlinear superposition formulas for the system of
differential equations (\ref{gr4}) which is connected with the
action of the Lie group ${\rm SO}(3,2)$ on space $M$.

\vskip0.5cm \noindent {\sf Comments:} Equations (\ref{sf6.4})
imply that four solutions are not fully independent. For example,
if we know three solutions ${\bf x}_i$, ${\bf Y}_i$, $i=1,2,3$,
and from the fourth we know $x_3^{(4)}$, we can obtain
$x_1^{(4)}$ and $x_2^{(4)}$ from (\ref{sf6.4}). This is a
consequence of the fact that the reconstruction of action of the
group requires only $10$ independent functions.

\vskip12pt

\begin{center}
{\bf 5. Conclusions}
\end{center}

The main results of this paper are the following:

1. We constructed systems of first--order ordinary differential
equations that arise from the infinitesimal action
of the local Lie group ${\rm SO}(n+1,n)$ on the factor--space
$M={\rm SO}(n+1,n)/P$, where $P$ is one of the maximal parabolic
subgroups of ${\rm SO}(n+1,n)$. These systems allow a superposition
formula.

2. We found any set of invariants for these systems which are
expressed in terms the solutions. These are invariants of
matrices $\Omega_{ik}$. The matrices $\Omega_{ik}$ play, in our
case, a similar role as the matrix anharmonic ratio for projective
matrix Riccati equations in \cite{JMP1}.

3. In the case of ${\rm SO}(3,2)$, we proved that, from these invariant
it is possible to find the general solution of our system on the
basis of four particular solutions. Therefore, in this case, these
set of invariants gives implicit nonlinear superposition formula.
It is necessary to note, that, even though the local groups
${\rm SO}(3,2)$ and ${\rm SP}(4,\R)$ are isomorphic, our system
of differential equations differs from the one studied in \cite{JMP1}
for the Lie group ${\rm SP}(4,\R)$, because we use another maximal
parabolic subgroup for constructing the space $M$.

\end{document}